\documentclass[aps, prb, twocolumn, floatfix, superscriptaddress, reprint, 10pt]{revtex4-1}  
\usepackage{graphicx}
\usepackage[usenames]{color}
\usepackage{amsmath}
\usepackage{amssymb}
\newcommand{\ang}{\mbox{\normalfont\AA}}

\begin{document}

\title{Generalized convex hull construction for materials discovery}

\author{Andrea Anelli}
\affiliation{Laboratory of Computational Science and Modeling, IMX, \'Ecole Polytechnique F\'ed\'erale de Lausanne, 1015 Lausanne, Switzerland}
\author{Edgar A Engel}
\email{E-mail address: edgar.engel@epfl.ch}
\affiliation{Laboratory of Computational Science and Modeling, IMX, \'Ecole Polytechnique F\'ed\'erale de Lausanne, 1015 Lausanne, Switzerland}
\author{Chris J. Pickard}
\affiliation{Department of Materials Science and Metallurgy, 27 Charles Babbage Road, Cambridge CB3 0FS, United Kingdom}
\affiliation{Advanced Institute for Materials Research, Tohoku University, 2-1-1 Katahira, Aoba, Sendai 980-8577, Japan}
\author{Michele Ceriotti}
\affiliation{Laboratory of Computational Science and Modeling, IMX, \'Ecole Polytechnique F\'ed\'erale de Lausanne, 1015 Lausanne, Switzerland}

\begin{abstract}
Searching for novel materials involves identifying potential candidates and selecting those that have desirable properties and facile synthesis. 
It is relatively easy to generate large numbers of potential candidates, for instance by computational searches or elemental substitution. The identification of synthesizable compounds, however, is a needle-in-a-haystack problem.
Conventionally, the screening is based on a convex hull construction, which identifies structures stabilized by a particular thermodynamic constraint, such as pressure, chosen based on prior experimental evidence or intuition.
We introduce a generalized convex hull framework that instead relies on data-driven coordinates, and represents the full structural diversity of the candidate compounds in an unbiased way. Its probabilistic construction addresses the inevitable uncertainty in input structure data and provides a superior measure of stability compared to the input (free) energies, that can for instance also be used to assist experimental crystal structure determination.
It efficiently identifies candidates with high probabilities of being synthesizable and suggests the relevant experimentally realizable constraints, thereby providing a much needed starting point for the determination of viable synthetic pathways.
\end{abstract}

\maketitle

\section{Introduction}
\label{sec:Introduction}

The aspiration of computational materials science is to autonomously predict structures with desirable properties and to design technologically relevant materials. 
This poses three main challenges: (i) comprehensively surveying the high-dimensional configuration space describing all possible structures, (ii) identifying experimentally and technologically relevant structures from a virtually infinite zoo of possible (meta)stable configurations, and (iii) designing experimental protocols to synthesize the structures of interest.

\begin{figure*}[tbh]
  \begin{minipage}[c]{1.0\textwidth}
    \includegraphics[width=\textwidth]{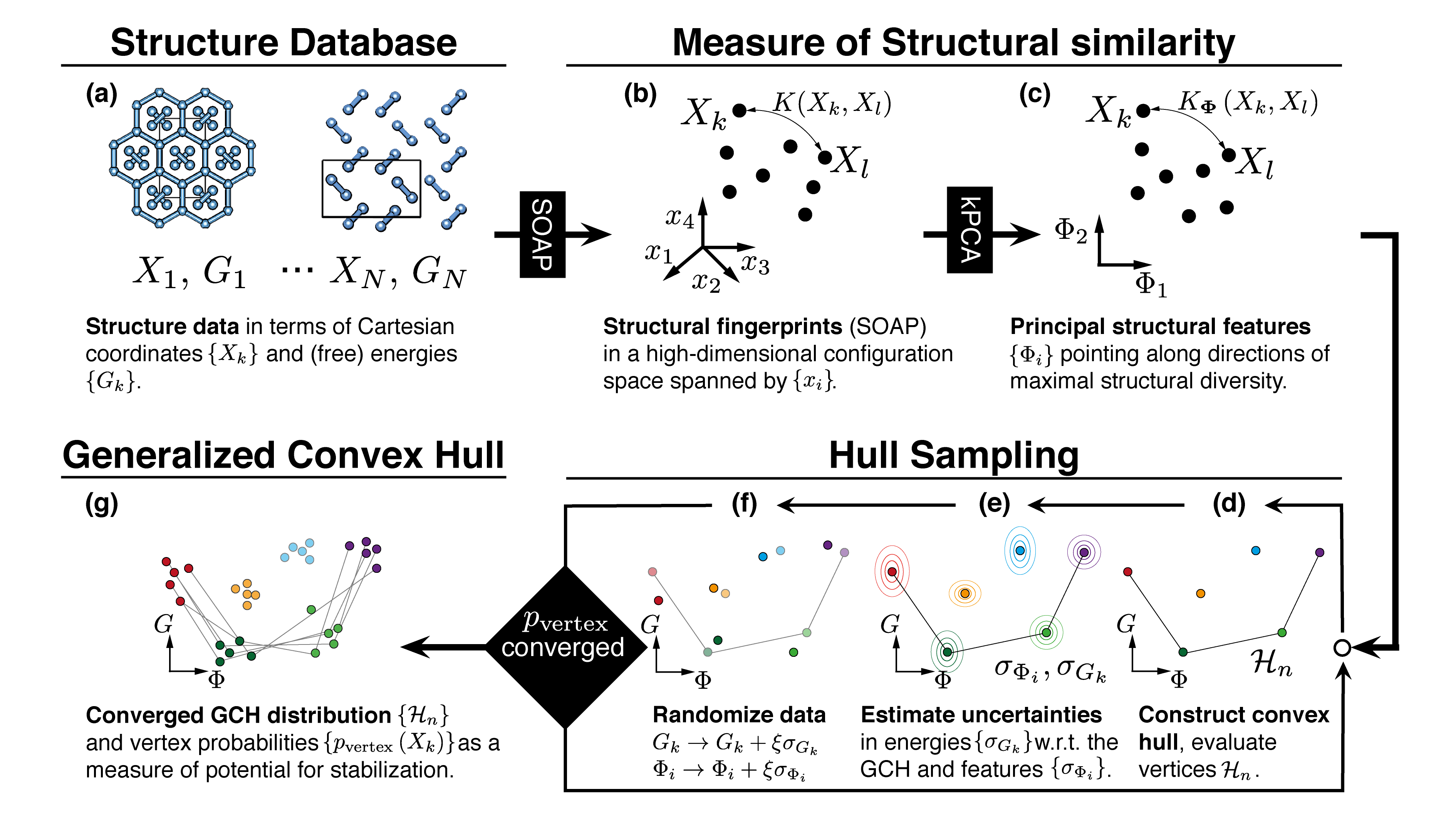}
  \end{minipage}\hfill
    \caption{
       Schematic representation of the GCH framework. 
    $X_k$ denotes structure $k$ with (free) energy $G_k$ and the associated (SOAP) structural descriptors $x_i(X_k)$ and (PCA) principal features $\boldsymbol{\Phi}(X_k) = \{\Phi_i (X_k)\}$. $\mathcal{H}{}_n$, $\sigma_{G_k}$, and $\sigma_{\Phi_i}$ denote the $n$th convex hull, the uncertainty in the (free) energy of $X_k$ with respect to the current convex hull, and the uncertainty in $\Phi_i$, respectively.
    $\xi$ are normally distributed random numbers and $p_{\mathrm{vertex}}(X_k)$ denotes the fraction of the sampled hulls for which $X_k \in \cal{H}$ (as a measure of the stabilizability of $X_k$).}
    \label{fig:scheme}
\end{figure*}

Numerous applications such as Refs.~\cite{pickard_2012,azadi_2014,drummond_2015,errea_2015,drozdov_2015,mayo_2016,monserrat_2016} demonstrate how configuration spaces can be explored effectively by combining atomistic calculations with various structure searching techniques \cite{pickard_2006,glass_2006,amsler_2010,yu_2011,zhu_2012,ong_2013,reilly_2016}, despite the exponential increase in the number of computationally (meta)stable structures with system size \cite{stillinger_1999}.
Meanwhile, the computational effort involved in (i) mapping phase diagrams using extensive Gibbs free energy calculations and (ii) determining possible synthetic pathways using methods such as forward flux sampling \cite{allen_2006} and enhanced sampling metadynamics  \cite{quigley_2009,giberti_2015} prevents bulk calculations for large numbers of locally stable structures. 
One of the key steps on the path to computational materials design is thus the reliable identification of the manageably small number of compounds stabilized by diverse thermodynamic conditions, given that geometries and relative stabilities are only available for one particular set of conditions. 

In the absence of kinetic effects \cite{ceder_2011} 
a convex hull (CH) construction can be used to identify structures and compounds that are stable with respect to decomposition into two or more parent structures at fixed thermodynamic conditions. 
For instance, consider the volume-based CH. If two structures $A$ and $B$ with molar volumes $V(A)$ and $V(B)$ and free energies $G(A)$ and $G(B)$ are part of the hull, then any structure $C$ with molar volume $V(A)<V(C)<V(B)$ and a free energy that lies above the line joining $A$ and $B$ on a $V-G$ plot will spontaneously decompose at constant volume into a mixture of $A$ and $B$ (see Fig.~\ref{fig:scheme} (d), taking $\phi = V$). 

CH constructions have proven useful in numerous structure searching applications such as Refs.~\cite{errea_2015,drozdov_2015,mayo_2016,pickard_2012,azadi_2014,drummond_2015,monserrat_2016}.
However, the conventional form has some crucial limitations.
The choice of one particular feature, such as molar volume, on which the CH is constructed, relies on experimental evidence or preconceived notions of which thermodynamic constraints may stabilize structures of interest. 
It limits which stabilizable structures are identified, and is generally insufficient to explore the structural diversity that can be accessed experimentally through complex thermodynamic constraints such as pressure, composition, doping with guest molecules, substitution of portions of organic compounds, electric or magnetic fields, etc. (for instance, see Ref.~\cite{pulido_2017}).
Furthermore, the conventional CH construction neglects inevitable inaccuracies in the computed (free) energies and geometries, which render the CH probabilistic in nature.

While the identification of experimentally-synthesizable compounds is the focus of this work, the generalized CH framework proposed in the following also translates (at negligible computational cost) input energies into a far better measure of structural stability, namely the energy relative to the GCH. The latter can be used in place of bare energies in diverse applications, such as experimental crystal structure determination protocols or as the fitness function driving structure searches \textit{in situ}.
\vspace{0.25cm}

\section{The Generalized Convex Hull}
\label{sec:GCH}

To overcome the above limitations we introduce a probabilistic generalized CH (GCH) framework for evaluating the probabilities of  structures being stabilized by general thermodynamic constraints. A schematic representation of this framework is shown in Fig.~\ref{fig:scheme}. It (i) quantifies the uncertainty arising from the inevitable errors in the underlying energies and structures, and (ii) rests on geometric fingerprints $\boldsymbol{\Phi} = \{ \Phi_i \}$, which reflect the full structural diversity of the dataset.
While there is considerable freedom in choosing such fingerprints, they must exhibit an additive behavior, that guarantees that a macroscopic sample $X$, which is a phase-separated mixture of different components $X_k$ with molar fractions $w_k$, has a fingerprint $\boldsymbol{\Phi}(X)=\sum_{k} w_k\boldsymbol{\Phi}(X_k)$.
A simple way to guarantee that $\Phi_i$ fulfills this requirement is to choose descriptors that are
consistent with an atom-based decomposition, $\boldsymbol{\Phi}(X)=\sum_{\mathcal{X}\in X}\boldsymbol{\phi}(\mathcal{X})/N_X$. 
Here $\boldsymbol{\phi}(\mathcal{X})$ are the fingerprints of the $N_X$ atom-centered, local environments $\mathcal{X}$ within the structure $X$.
Additivity ensures that any structure with features inside a convex region of $D$-dimensional feature-space can be decomposed into a phase-separated mixture of the $D+1$ vertices of the convex region, without changing the corresponding $D$ features of the fingerprint describing the system (although the resultant fingerprint may differ in the remaining features).
By considering the molar free energy as a function of a set of $D$ features $\Phi_i$, one can thus generalize the CH construction to identify the structures that are stable with respect to decomposition subject to the abstract ``thermodynamic constraint'' defined by a given set of $D$ features.
\vspace{0.25cm}

\textbf{Data-driven structure fingerprints.}
For a given dataset $\left\{X_k\right\}$ we extract a small set of key data-driven features that captures most of its structural diversity by performing a kernel principal component analysis (KPCA) on a kernel measure of similarity $K(X_k,X_l)$ between pairs of structures $X_k$ and $X_l$. That is, we compute the eigenvalues $\lambda_i$ and eigenvectors  ${\bf u}^i$ of the kernel matrix, $K_{kl} = K(X_k, X_l)$, and evaluate the features of a structure $X_k$ as
\begin{equation}
\Phi_i(X_k) = \sum_l u^i_l \sqrt{\lambda_i} K_{kl}.
\label{eq:features1}
\end{equation}
These features are additive, provided that $K(X_k,X_l)=\sum_{\mathcal{X}_k\in X_k,\mathcal{X}_l \in X_l} k(\mathcal{X}_k,\mathcal{X}_l)/N_{X_k} N_{X_l}$, where $k(\mathcal{X}_k,\mathcal{X}_l)$ is a kernel measure of similarity between pairs of local environments $\mathcal{X}_k$ and $\mathcal{X}_l$.

In practice, we use the smooth overlap of atomic positions (SOAP) kernel, which is constructed around atom-centered, local environments and thus additive, but is otherwise general and agnostic, and can be applied seamlessly to different kinds of materials~\cite{szlachta_2014,deringer_2017,bartok_2017,musil_2018}. Crucially, SOAP fingerprints have proven reliable for both energy regression~\cite{bartok_2017} and structure classification~\cite{de_2016} for the systems discussed in the following. 
SOAP describes an atomic environment $\mathcal{X}$ as a sum of atom-centered Gaussians
\begin{equation}
\langle {\alpha \bf r} | \mathcal{X} \rangle = \sum_{j \in \mathcal{X},\alpha} \exp\left( - \frac{({\bf x}_j - {\bf r})^2}{2\sigma^2} \right) \, ,
\label{eq:features2}
\end{equation}
where ${\bf x}_j$ are the Cartesian coordinates of the atoms of chemical identity $\alpha$ (H, O, C, $\ldots$) within a radial cutoff $r_c$.  $\sigma$ specifies the width associated with each atomic probability distribution.
The (rotationally averaged) power spectrum of the expansion of an environment $\mathcal{X}$ on a basis of radial functions $R_n(r)$ and spherical harmonics $Y_{lm}({\bf \hat{r}})$,
\begin{equation}
\begin{split}
\langle \alpha n \alpha'n'l | \mathcal{X} \rangle &\propto \sum_m \langle \alpha nlm | \mathcal{X} \rangle \langle \mathcal{X} | \alpha' n'lm \rangle \, ,\\
\langle \alpha nlm | \mathcal{X} \rangle &= \int \mathrm{d}{\bf r} \, R_n(r) Y_{lm}({\bf \hat{r}}) \langle {\alpha \bf r} | \mathcal{X} \rangle \, , 
\end{split}
\label{eq:features3}
\end{equation}
provides the representation to define the environmental kernels 
\begin{equation}
k(\mathcal{X},\mathcal{X}')=\left[\sum_{\alpha n \alpha'n'l} \langle \alpha n \alpha'n'l | \mathcal{X} \rangle \langle \alpha n \alpha'n'l | \mathcal{X}' \rangle\right]^2 \, .
\label{eq:features4}
\end{equation}
Loosely speaking, the resultant KPCA features $\Phi_i(X_k)$ are orthonormal measures of the similarity of the structure $X_k$ to a particular combination of all structures in the dataset, dominated by the structurally most distinct configurations.
\vspace{0.25cm}

\begin{figure}[tb]
    \centering
    \includegraphics[width=0.4\textwidth]{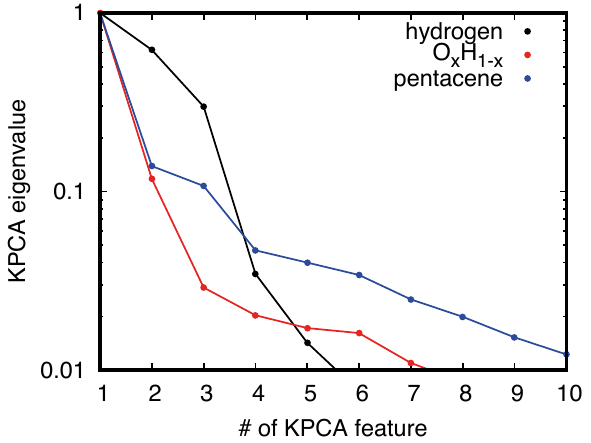}
    \caption{KPCA eigenvalues for the applications we discuss in this work, namely: hydrogen (black), H$_x$O$_{1-x}$ (red), and pentacene (blue),  obtained from SOAP similarity kernels with $r_c = 2\,\ang$, $5\,\ang$, and $5\,\ang$, respectively.}
    \label{fig:kpca_eigenv_spectrum}
\end{figure}

\textbf{Feature selection and interpretation.}
The abstract nature of these KPCA features begs the question of (i) how to identify which among them have the potential to stabilize structures and should thus be included in the GCH construction, and (ii) how to relate them to experimentally realizable conditions. 
When no prior knowledge of the system is available the KPCA eigenvalue spectrum provides indication of the maximum intrinsic dimensionality (Fig.~\ref{fig:kpca_eigenv_spectrum}) of the structure data at hand~\cite{fukunaga_1971}. It can thus be used to choose the dimensionality of the GCH such that the full structural diversity of the dataset is explored. 
Even in this worst case scenario, the resultant pool of candidates is typically orders of magnitude smaller than the underlying structure database, rendering it possible to further investigate the relations between the features of the candidates and physical observables (or thermodynamic constraints) such as density, composition, etc.
This can not only help to translate abstract structural features into practically realizable synthetic protocols, but also to refine the selection of features on which the GCH is constructed \textit{a posteriori} to those which couple strongly to experimentally realizable conditions and thus have the greatest potential for stabilizing structures.
\vspace{0.25cm}

\begin{figure*}[tbh]
\centering
\includegraphics[width=1.0\textwidth]{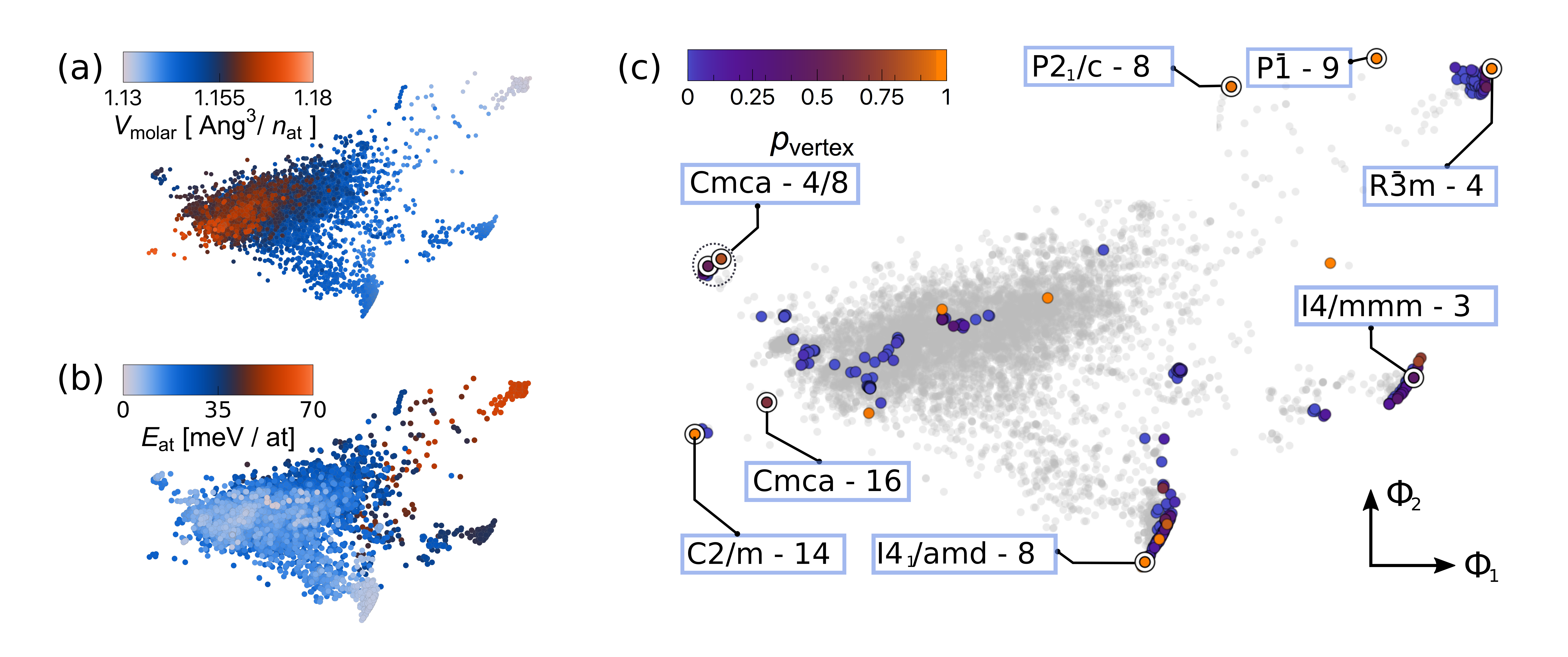}
\caption{Maps of 7,594 hydrogen structures spanned by the two dominant KPCA features, $\Phi_1$ and $\Phi_2$. Due to their abstract nature (Eqs.~(\ref{eq:features1}) to (\ref{eq:features4})) the numerical value of $\Phi_1$ and $\Phi_2$ is not shown. 
Each point corresponds to a structure in the dataset. The maps on the left are colored according to (a) molar volume and (b) molar energy. One can see the clear correlation between the KPCA coordinates, and structural and energetic properties. 
(c) The larger map highlights structures with non-negligible probability $p_{\textrm{vertex}}$ of being part of the GCH built on the first four KPCA features, which is represented as a color scale. Candidates surviving an additional ``coarse-graining'' step down to the point where all remaining structures have $p_\text{vertex}=1$ are labeled according to space group and number of atoms per unit cell.
By comparison with the map colored according to molar energy, one sees that the convex hull identifies clusters of configurations that are low in energy and/or extremal in structure.}
\label{fig:hydrogen_map}
\end{figure*}

\textbf{Probabilistic GCH and uncertainty quantification.} 
So far, the GCH framework neglects the inevitable uncertainties in (computed) free energies, lattice parameters and atomic positions, and therefore in the determination of the hull vertices, making it necessary to use rather arbitrary confidence regions around the hull, and to manually remove duplicate structures \cite{engel_2018}. We therefore propose a probabilistic extension in which the GCH probability distribution is sampled by constructing many possible convex hulls based on free energies and geometries, which have been randomized according to their respective uncertainties. 
In practice we take the typical model errors on the energies $\epsilon$ and Cartesian coordinates (for example, due the choice of density functional in density functional theory (DFT) calculations or the absence of quantum nuclear effects) to be known from experience or benchmarks.  
We estimate the resultant errors in the energies \textit{relative to the instantaneous hull}, $\sigma_{G_k}$, exploiting structural correlations to account for correlations between the errors in $\{G_k\}$. In particular we ensure that $\sigma_{G_k}$ vanishes for the vertex structures and any structure that is a phase-separated mixture of the vertices of its associated simplex (its ``parent phases''), while otherwise reflecting how different a given non-hull structure is from the parent phases.
The rationale is that the typical errors are not random, but correlate with the structural features.
Consider for instance a phase-separated mixture $X_k$ composed of molar fractions $w_{kl}$ of the parent phases $X_l$ with calculated energies $G_l + \epsilon_l$. Its calculated energy is identical to the corresponding combination of the energies of the parent phases, including their errors,  $\sum_{l} w_{kl} ( G_l + \epsilon_l)$. 
This is exactly the definition of the convex hull energy constructed on $ G_l + \epsilon_l$, so that the energy of $X_k$ relative to the hull will be zero regardless of the errors. Hence, $\sigma_{G_k}$ should vanish. 

Let us introduce a practical definition that satisfies this requirement. We estimate $\sigma_{G_k}$ as the fraction of the total error $\epsilon$ associated with the deviation of the features $\Phi_i(X_k)$ from the ideal interpolation in terms of the parent phases, $\Phi_i^\text{GCH}(X_k) \equiv \sum_{X_j \in {\cal H}} w_{kj} \Phi_i(X_j)$
\begin{equation}
\sigma_{G_k} = \epsilon \sqrt{\frac{1}{\sigma^2_G} \sum_{i=1} \left[ g_i
\left(\Phi_i(X_k) - \Phi_i^{\textrm{GCH}}\right)\right]^2} \, .
\end{equation}
Here $g_i$ is the energetic response to changes in $\Phi_i$, which we learn by ridge regression from a machine-learning model of $G_k$, and $\sigma^2_G$ is the variance of $G$ over the entire dataset.
Due to additivity, for a physical mixture, $\Phi_i(X_k)=\Phi_i^\text{GCH}(X_k)$ for all the features, including those that are not used for the GCH construction, which ensures that $\sigma_{G_k}=0$. 
On the contrary, for each point that is not a physical mixture of hull points, only the features used to build the GCH will coincide with $\Phi_i^\text{GCH}(X_k)$. In this case, $\sigma_{G_k}$ scales with the residual structural diversity that is not captured by the GCH coordinates. 
Note that the dependence of the uncertainties $\sigma_{G_k}$ on the instantaneous hull implies that the hull distribution must be sampled ``self-consistently''.

The randomization of the features $\Phi_i$ requires knowledge of how the uncertainty in the underlying atomic coordinates and lattice parameters (or ``structure parameters'') propagates to uncertainties in the features, $\sigma_{\Phi_i}$.
We estimate $\sigma_{\Phi_i}$ by randomizing the structure parameters of $n_r$ reference configurations $X_r$, that we choose by farthest point sampling.
In practice, we randomize each reference structure $n_s$ times, compute the features for the randomized structures $\Phi_i (X^s_r)$, and evaluate
\begin{equation*}
\sigma_{\Phi_i} = 
\sqrt{
\frac{1}{n_s n_r} \sum_{r}^{n_r}
\sum_{s}^{n_s}\left( \Phi_i (X^s_r) - \Phi_i (X_r) \right)^2 }\, .
\end{equation*}
After extensive sampling of the GCH distribution the rate with which each structure occurs as a vertex $p_{\textrm{vertex}}(X_k)$ roughly quantifies how trustworthy the identification of the structure $X_k$ as stabilizable is and its average distance from the hull provides a measure of its (meta)stability.
\vspace{0.25cm}

\textbf{Coarse graining of the GCH vertices.}
In cases where large numbers of very similar structures (for example owing to stacking faults or partial disorder) compete for stability each candidate exhibits a small individual probability of becoming stable. However, collectively such a cluster of structures represents a stable phase.
For convenience we reduce the full list of potential vertices to 
representatives of each cluster, that is, of each stable phase. 
These are identified by sequentially eliminating the $N$ lowest probability candidates with a cumulative probability $\sum_{k=1}^N p_{\textrm{vertex}}(X_k) < 1$ (which guarantees that no complete cluster of structures that constitutes one stabilizable structure gets eliminated entirely in one step) from the dataset and resampling the GCH for the thus reduced dataset. This procedure is repeated until the lowest $p_{\textrm{vertex}}(X_k)$ is above a set threshold of 0.5. 
This ``coarse-graining'' ensures that the surviving candidates correctly accumulate the probability of becoming stable associated with their respective clusters of similar structures.
Even though we only consider these marginal probabilities, the GCH directly samples the full hull distribution, which can further be used to investigate for instance which structures compete for stability.
\vspace{0.25cm}

\section{Applications}
\label{sec:Applications}
To demonstrate the power of the GCH framework, we apply it to four problems of increasing complexity, namely a database of hydrogen solid phases at terapascal pressure, a set of oxygen-hydrogen binary crystal structures, a subset of this database for which we demonstrate how a GCH can predict oxygen phases that are stabilized by magnetism, and a set of crystalline polymorphs of pentacene for which we investigate chemical substitutions and demonstrate the stability of the GCH to errors in the input energy data.
The respective structure databases are available as supplemental material.
\vspace{0.25cm}

\textbf{Hydrogen at gigapascal pressure.}
As a first test, we analyze 7,964 locally stable hydrogen structures from an \textit{ab initio} random structure search (AIRSS)~\cite{pickard_2012,pickard_2012_2} at 500\,GPa based on DFT geometry optimizations using the Perdew-Burke-Ernzerhof (PBE) functional~\cite{perdew_1996}, where extensive experimental and theoretical literature \cite{eremets_2011,mcmahon_2011,mcmahon_2012,eremets_2016,dalladay_2016} provides a detailed reference of stabilizable structures.
Fig.~\ref{fig:hydrogen_map} shows a representation of the GCH procedure when performed on the dominant two KPCA components resulting from a SOAP kernel ($r_c = 2 \ang$). It is clear that the principal components correlate strongly with the cohesive energy and the molar volume of the structures, and that the GCH procedure identifies configurations that are extremal in geometry and/or particularly favorable energetically. 
While this two-dimensional map provides for a more intuitive visualization, the KPCA eigenvalue spectrum  (see Fig.~\ref{fig:kpca_eigenv_spectrum})  suggests that the intrinsic dimensionality of the dataset is higher. We therefore consider for further analysis the GCH constructed on the top four components.
We identify 81 candidate structures, and successfully recover the high-pressure molecular $I4_1amd$ and atomic $R\bar{3}m$ phases of hydrogen, as well as analogs of the lower-pressure phases II to IV (a comparison between the structures and their lower-pressure analogs is given in the SI). The latter are not stable at the simulated conditions, so being able to find very similar structures among the candidates is a testament to the long-sightedness of AIRSS and the predictive power of the GCH. 
To achieve the same feat using a conventional energy-volume CH, structures up to around 8\,meV/atom above the CH have to be retained, leaving a disproportionately larger pool of more than 2,000 potentially stabilizable structures. 
\vspace{0.25cm}

\begin{figure}
(a)\hspace{7cm}\phantom{a}\\
\includegraphics[width=0.45\textwidth]{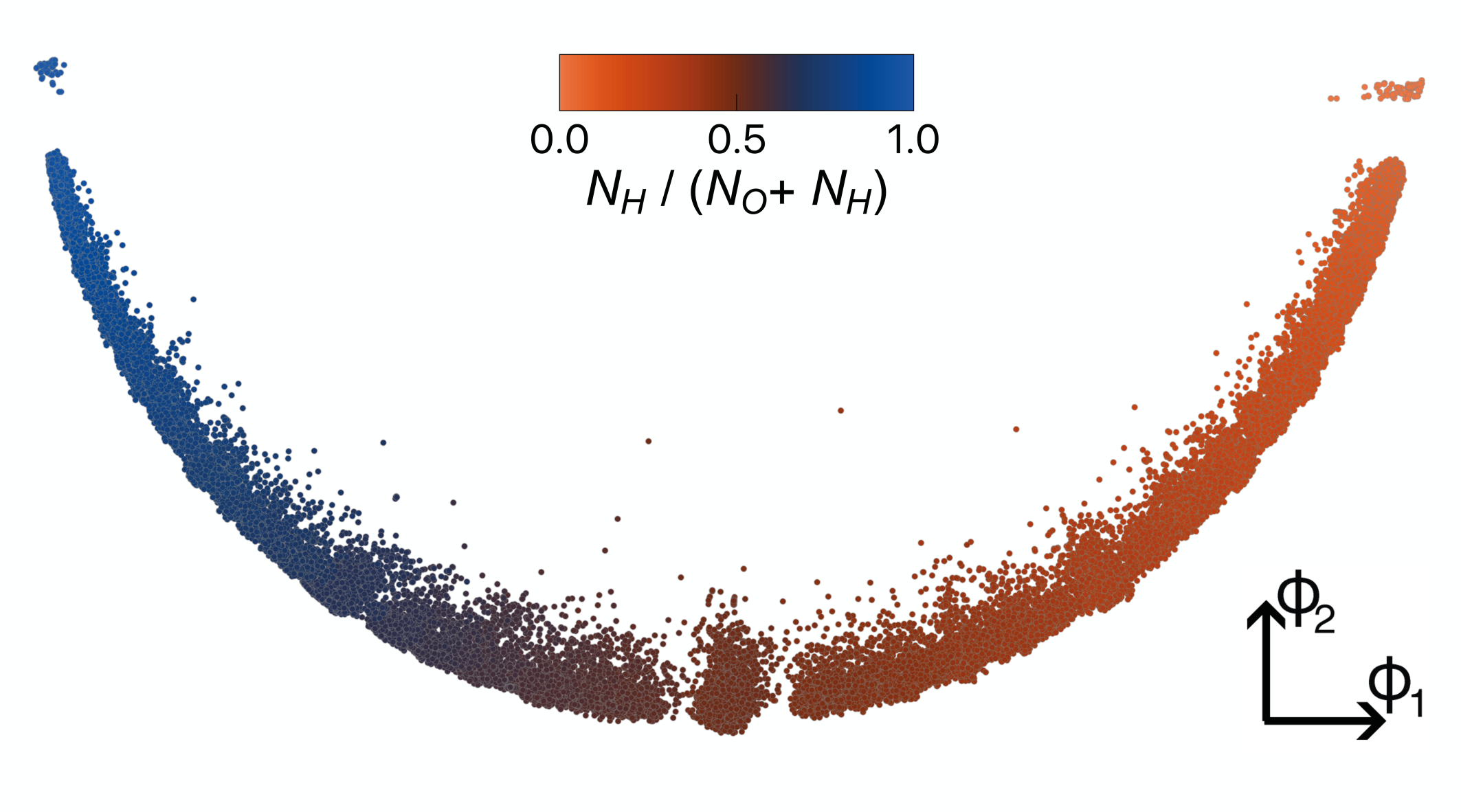}\\
(b)\hspace{7cm}\phantom{a}\\
\includegraphics[width=0.45\textwidth]{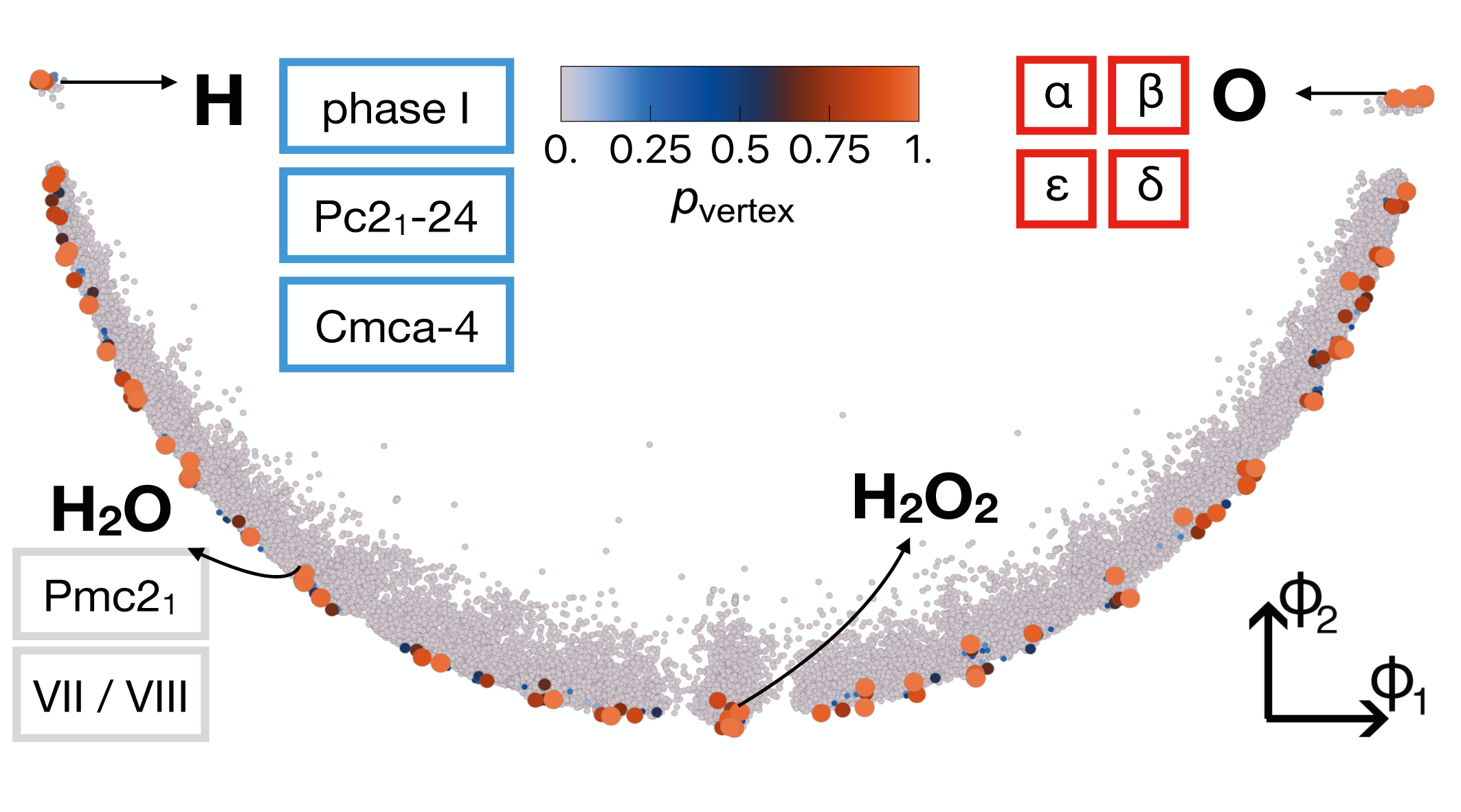}
\caption{Map of 51,376 H$_x$O$_{1-x}$ structures spanned by the two dominant KPCA features, $\Phi_1$ and $\Phi_2$. 
The structures are colored according to (a) composition, and (b) their probability, $p_{\textrm{vertex}}$, of constituting a vertex of the CH of $E(\Phi_1,\Phi_2)$. 
The positions of experimentally-confirmed and proposed hydrogen, ice, hydrogen peroxide, and oxygen structures are highlighted. Proposed structures are labeled according to their symmetry group.}
\label{fig:ho_map}
\end{figure}

\textbf{Oxygen-Hydrogen binary compounds.}
The next level of complexity in computational materials discovery involves the modeling of multi-component systems: in the case of 51,376 locally stable H$_x$O$_{1-x}$ configurations from an \textit{ab initio} random structure search (AIRSS)~\cite{pickard_2012,pickard_2012_2} at 20\,GPa (again based on DFT geometry optimizations using the PBE functional~\cite{perdew_1996}) the GCH framework must resolve the most stable stoichiometries, while at the same time recovering various hydrogen, ice and oxygen phases. 
The KPCA eigenvalues based on a SOAP kernel ($r_c = 5 \ang$) decay by more than an order of magnitude after the first feature (see Fig.~\ref{fig:kpca_eigenv_spectrum}).
This reflects the dominant role of composition in determining structural diversity and forecloses the identification of the first KPCA feature with composition (see Fig.~\ref{fig:ho_map} (a)).
Along this principal axis, one identifies the expected stable oxygen, hydrogen, and ice structures, but also crystalline hydrogen peroxide, ice phases with different fractions of intercalated hydrogen molecules and crystalline molecular hydrogen and oxygen phases with guest water molecules.
The latter are unstable in the absence of other stabilizing fields as highlighted by an energy-composition CH construction. Their stability on the GCH arises because the first KPCA feature (while predominantly describing composition) also measures molar volume as an additional stabilizing factor.
When constructed on the first two KPCA features the GCH framework identifies 171 stabilizable structures, differing in both stoichiometry and geometry (see Fig.~\ref{fig:ho_map} (b)). Among nine hydrogen structures are phase I, the $Pc2_1$-24 candidate for phase II, and the $Cmca$-4 candidate for phase IV \cite{drummond_2015}. 
Reassuringly, the three ice phases include the experimentally stable ice VII/VIII and the $Pmc2_1$ high-pressure candidate phase of Hermann \textit{et al.} \cite{hermann_2012}. 
\vspace{0.25cm}

\begin{figure}
\includegraphics[width=0.45\textwidth]{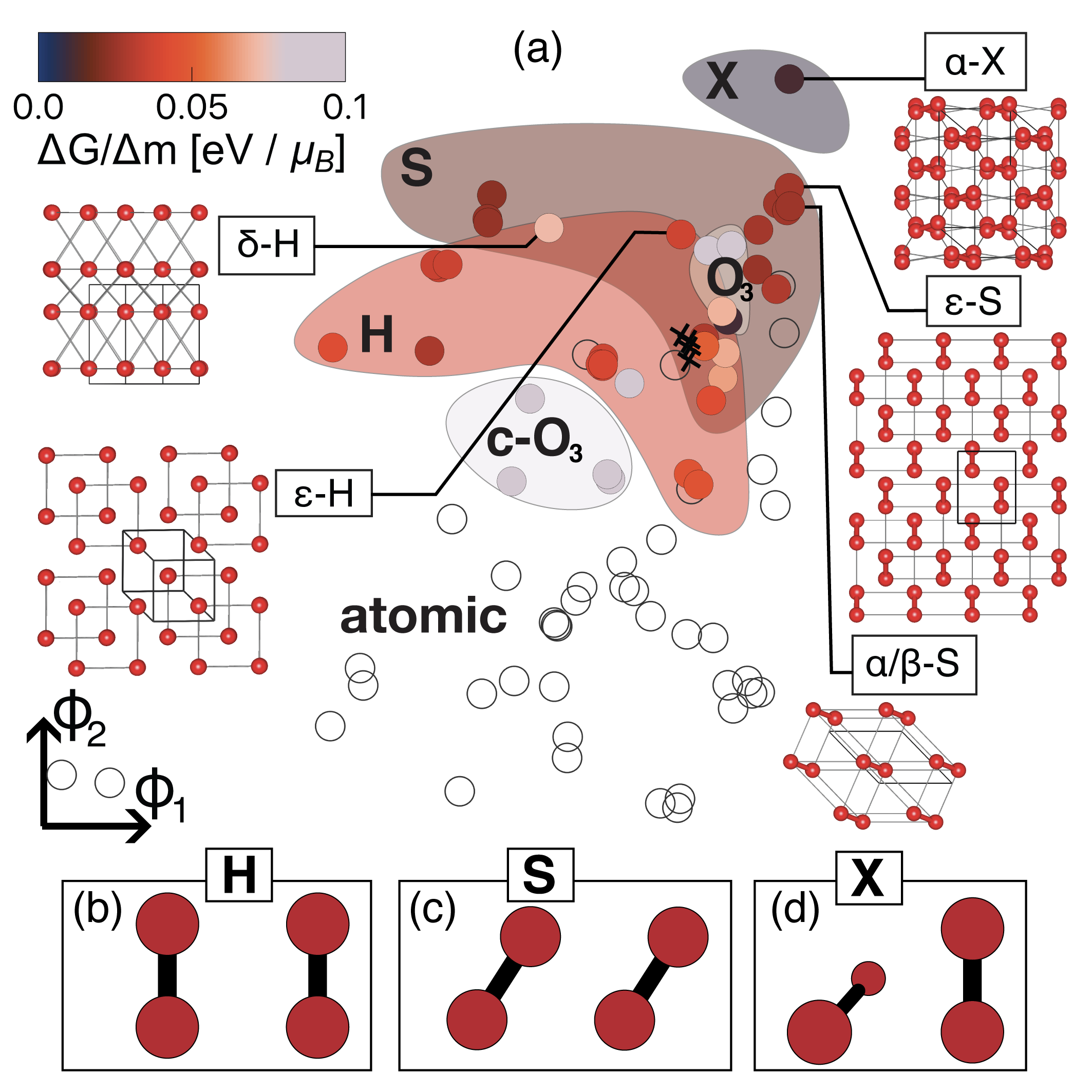}
\caption{PCA projection of the subset of 84 pure oxygen structures onto $\Phi_1$ and $\Phi_2$ as obtained for the full dataset of 51,376 H$_x$O$_{1-x}$ structures. 
Diamagnetic molecular structures (solid circles) are colored according to $\Delta G / \Delta m$.
Atomic and ferromagnetic molecular structures are shown as open circles and crosses, respectively.
The shaded regions highlight molecular structures in the H, S, and X configurations ((b) to (d)), and are colored according to the respective mean values of $\Delta G / \Delta m$. 
This highlights the correlation between $\Phi_{1,2}$, molecular tilts, and energetic response to magnetization $\Delta G / \Delta m$ as a proxy of the potential for stabilization by magnetic fields.}
\label{fig:oxygen_map}
\end{figure}

\textbf{Magnetically-stabilized phases of oxygen.}
The six oxygen structures deserve a more detailed discussion, as they demonstrate that the GCH is capable of revealing subtle mechanisms of stabilization, which have barely been touched upon in literature, such as the stabilization of unconventional molecular oxygen phases by external magnetic fields.
Using the nomenclature introduced in Refs.~\citenum{defotis_1981,hemert_1983,kitaura_2002}, the six oxygen structures include the conventional $\alpha$/$\beta$ and $\varepsilon$~\cite{freiman_2004} phases, in which the O$_2$ molecules align in the so called ``H''-state (Fig.~\ref{fig:oxygen_map} (b)).  
The GCH further detects $\alpha$/$\beta$ and $\delta$ phases with uniformly-tilted O$_2$ molecules (``S'' state, Fig.~\ref{fig:oxygen_map} (c)) and an $\alpha$ phase, in which the molecules display an alternating tilt pattern (``X'' state, Fig.~\ref{fig:oxygen_map} (d)).
Experimental evidence suggests that these may be stabilized by strong magnetic fields \cite{kitaura_2002,nomura_2014}, which we further substantiate using spin-polarized DFT calculations using QUANTUM ESPRESSO \cite{giannozzi_2009} (see Fig.~\ref{fig:oxygen_map} and SI Fig.~S5). This demonstrates (i) that structural features do indeed correlate with subtle responses to manipulations of the electronic structure of a configuration and (ii) how one can verify the coupling between abstract structural coordinates and experimentally realizable thermodynamic constraints.
\vspace{0.25cm}

\begin{figure}
\includegraphics[width=0.45\textwidth]{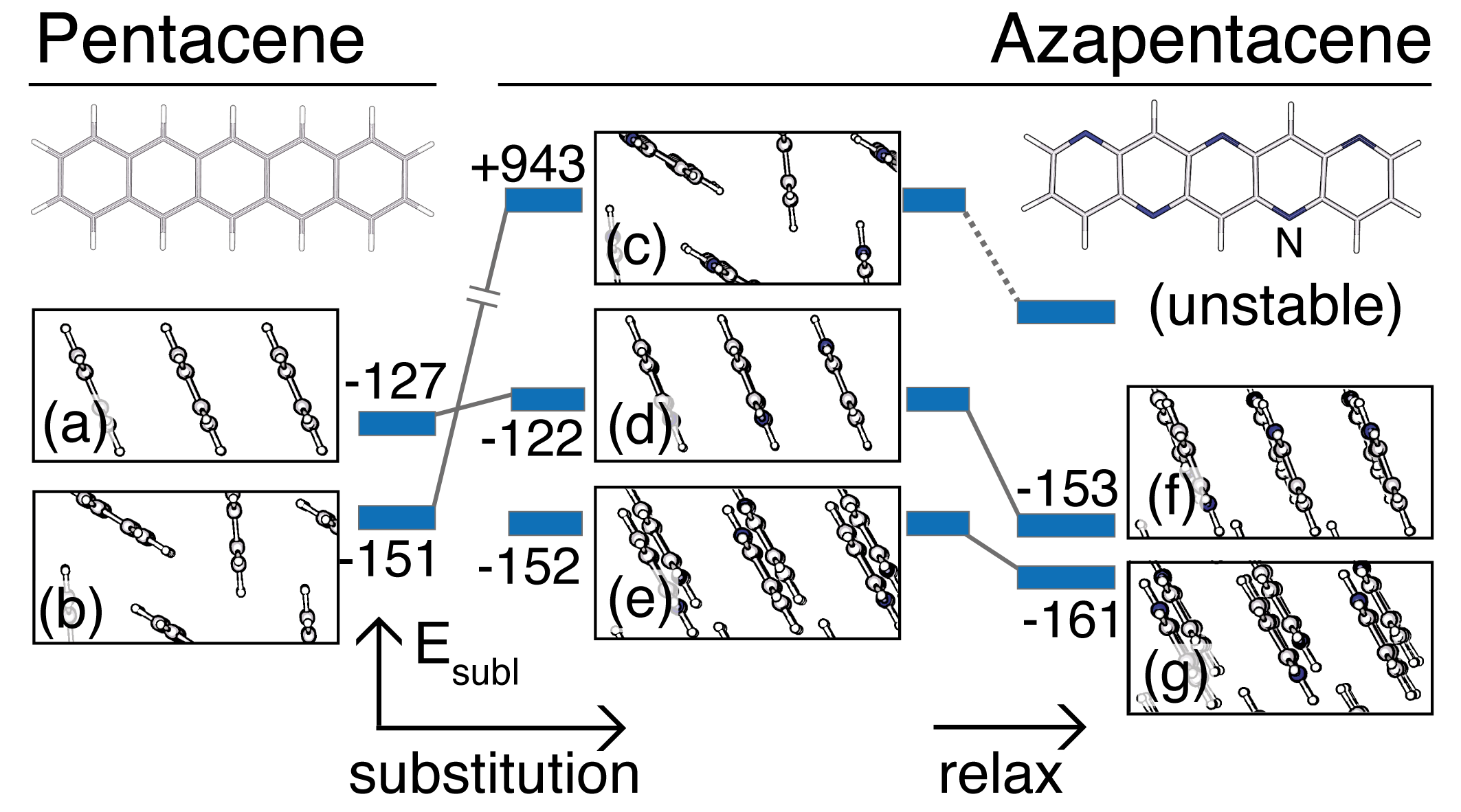}
\caption{Sublimation energies, $E_{\textrm{subl}}$, of different pentacene configurations in kJ/mol before (left) and after 5A nitrogen substitution (center), and after subsequent geometry optimization (right). 
(a) is among the most unstable pentacene configurations in the dataset. (e) is the most stable 5A substituted azapentacene configuration among 594 configurations from an independent structure search~\cite{campbell_2017}.
The $E_{\textrm{subl}}$ computed for the Campbell bulk phase (b) of 151.019\,kJ/mol agrees with the experimental values of $154.5$~\cite{kruif_1980} and $156.9\pm13.6$\,kJ/mol~\cite{oja_1998} to within the errors.
}
\label{fig:pentacene}
\end{figure}

\textbf{Nitrogen substitution in pentacene.}
As a final example, we analyze a database of 564 locally stable arrangements of pentacene molecules. This application beyond high-pressure physics demonstrates how the GCH can suggest suitable starting points for studies of chemical substitution.
The configurations were obtained by a systematic structure search~\cite{campbell_2017}, based on rigid, DFT-optimized molecular units interacting via the W99 force-field~\cite{williams_1999}. 
In Ref.~\cite{campbell_2017}, this structure search is accompanied by independent searches for 5A (see Fig.~\ref{fig:pentacene}) and 5B nitrogen-substituted molecules, which are required because the stability of a given molecule is rarely a good predictor of the behavior of its substituted counterparts~\cite{giangreco_2017}.

We first perform a KPCA of the pentacene dataset using the same SOAP kernel ($r_c = 5 \ang$ and $\sigma = 0.3 \ang$) which has previously proven suitable for energy regressions~\cite{musil_2018}.
Alongside conventional, energetically favorable herringbone configurations, such as the Campbell bulk phase \cite{campbell_1961}, the GCH constructed on the two dominant KPCA features identifies five energetically unfavorable configurations with planar, colinear arrangements of molecules as stabilizable. 
Whereas nitrogen substitution of the global minimum pentacene configuration leads to a high-energy, unstable structure, several of the GCH vertices that are much higher in energy, which would therefore be discarded in a conventional analysis, retain their geometry upon nitrogen substitution and relaxation (see Fig.~\ref{fig:pentacene}).
Moreover, they exhibit competitive energies compared to the most stable 5A substituted configuration.
The GCH framework has thus effectively identified pentacene configurations with potential for stabilization by nitrogen substitution.
\vspace{0.25cm}

\begin{table}[t]
    \centering
    \begin{tabular}{lcc}
    \hline \hline
    & RMSE($\{E_k^{\textrm{W99}}\}$) & $\tilde{d}$ \\
    \hline
    $E$-$\rho$ CH   & 0.22  & 0.0139 \\
    d-GCH (1D)      & 0.11  & 0.0227 \\
    GCH (1D)        & 0.10  & 0.0168 \\
    GCH (1D) cg     &       & 0.0046 \\
    d-GCH (3D)      & 0.07  & 0.0087 \\
    GCH (3D)        & 0.07  & 0.0066 \\
    GCH (3D) cg     &       & 0.0009 \\
    \hline\hline
    \end{tabular}
    \caption{Sensitivity analysis of the (conventional) energy-density ($E$-$\rho$ CH) hull, and  deterministic (d-GCH), and probabilistic hulls (GCH) constructed on the first (1D) and first three (3D) KPCA features (before and after coarse-graining (cg)). Different metrics of the similarity of different CH constructions are evaluated on the basis of W99 and DFT sublimation energies for the 564 pentacene configurations from Ref.~\cite{campbell_2017}: (i) The RMSE in kJ/mol in the W99 convex hull energies $\{E_k^{\textrm{W99}}\}$ compared to ``reference'' DFT convex hull energies $\{E_k^{\textrm{DFT}}\}$ (for the full dataset), and (ii) the distance $\tilde{d}$ between the W99 and DFT based hulls as defined in Eq.~(\ref{eq:dhull}).
    \label{tab:sensitivity}}
\end{table}
\textbf{Sensitivity to errors in energetics.} 
The probabilistic sampling of the GCH does not only provide a robust strategy to eliminate redundant structures and for uncertainty quantification.
It also significantly reduces the sensitivity to errors in input energies compared to conventional deterministic CH constructions. 
To assess how sensitive different CH constructions are with respect to the details of the input energies, we calculate DFT sublimation energies using QUANTUM ESPRESSO \cite{giannozzi_2009} with the PBE functional and a Grimme-D2 dispersion correction~\footnote{We use a plane-wave energy cutoff of 100 Rydberg, a Monkhorst-Pack $\mathbf{k}$-point grid \cite{monkhorst_1976} spacing of less than $2\pi \times 0.07\ \ang^{-1}$, and the ultrasoft C.pbe-n-kjpaw\_psl.0.1.UPF, H.pbe-kjpaw\_psl.0.1.UPF, and N.pbe-n-kjpaw\_psl.0.1.UPF pseudopotentials from \textit{http://www.quantum-espresso.org}} for all 564 pentacene configurations for comparison with those obtained from the W99 force-field.
The DFT and W99 sublimation energies exhibit substantial differences (resulting in a root-mean-square error (RMSE) with respect to each other of 0.15\,kJ/mol after subtracting the respective averages), including a different global energy minimum structure.
As shown in Table~\ref{tab:sensitivity}, computing energies relative to the convex hull $E^{\textrm{DFT}/\textrm{W99}}_k = G_k^{\textrm{DFT}/\textrm{W99}} - \sum_l w_{kl}^{\textrm{DFT}/\textrm{W99}} G_l^{\textrm{DFT}/\textrm{W99}}$
reduces dramatically the discrepancy. This is a consequence of the fact that energy errors are correlated, which we also exploit in our probabilistic hull construction. 

The set of structures that are tagged as ``synthesizable'' is perhaps even more important than the estimate of the instability of the other candidates. 
Since different, structurally very similar configurations, for example only differing in proton or stacking (dis-)order, can be equivalently valid representatives of the same (stabilizable) phase, one cannot simply compare the indices of the structures identified as vertices.
To determine whether two hulls ${\cal H}_{\textrm{DFT}}$ and ${\cal H}_{\textrm{W99}}$ constructed on the basis of DFT and W99 energies, $\{ G_k^{\textrm{DFT}}\}$ and $\{ G_k^{\textrm{W99}}\}$, respectively, contain structurally similar vertices, we define a ``distance'' between hulls
as the mean minimum Euclidean distance between their respective vertices
\begin{equation}
\begin{split}
\tilde{d} 
&= \frac{1}{2} \left( d_{\textrm{DFT}}^{\textrm{W99}} + d_{\textrm{W99}}^{\textrm{DFT}} \right) 
\\
d_{\textrm{DFT}}^{\textrm{W99}} 
&= \sqrt{ \frac{1}{N_{\textrm{DFT}}}\sum_{X_i \in {\cal H}_{\textrm{DFT}}} \min_{X_j \in {\cal H}_{\textrm{W99}}} \left| \boldsymbol{\Phi}(X_i)-\boldsymbol{\Phi}(X_j) \right|^2 } \, .
\end{split}
\label{eq:dhull}
\end{equation}

The results of this analysis, shown in Table~\ref{tab:sensitivity} confirm that the GCH construction reduces the sensitivity of both the vertex selection and the measure of stability compared to a conventional construction. 
Increasing the dimensionality of the fingerprint space on which the hull is constructed, sampling probabilistically different realizations of the hull, and eliminating redundant structures in the database, all lead to a more robust determination of stabilizable structures that should be considered for further theoretical or experimental investigation. 
\vspace{0.25cm}

\section{Conclusions}
\label{sec:Conclusions}
These examples clearly evidence the wide spectrum of thermodynamic constraints which can be rationalized using the GCH framework and serve to showcase the remarkable versatility and transferability of the GCH framework.
The construction is only weakly dependent on the details of the kernel, and its probabilistic nature renders it robust to errors in the determination of the (free)-energies of different phases, which is very important given the harsh compromises one has to make between the accuracy and thoroughness of high-throughput structure searches.
Moreover, it is capable of eliminating redundant configurations in a physically meaningful way and of providing estimates of stability regimes in terms of experimentally realizable thermodynamic constraints.
The GCH framework provides a robust, data-driven, method- and error-insensitive evolution of the convex hull construction, one of the most essential tools to predict and rationalize the stability of materials, and to identify experimentally stabilizable structures among large numbers of locally stable configurations. 
\vspace{0.25cm}

%
\textit{Acknowledgements.}
M.C., A.A. and E.A.E. were supported by the European Research Council under the European Union's Horizon 2020 research and innovation programme (grant agreement no. 677013-HBMAP). 
C.J.P. is supported by the Royal Society through a Royal Society Wolfson Research Merit award. 
Calculations were performed on the Archer facility of the United Kingdom's national high-performance computing service (for which access was obtained via the UKCP consortium [EP/P022596/1]).
We would like to thank G.M. Day and J. Yang for sharing the W99 optimized configurations of pentacene and 5A crystals, and for insightful discussion.

%
%

\end{document}